\documentclass[sigconf]{acmart}
\usepackage{multirow}
\usepackage[normalem]{ulem}
\useunder{\uline}{\ul}{}
\usepackage{float}
\usepackage{graphicx}
\usepackage{marvosym}
\AtBeginDocument{%
  }

\setcopyright{acmcopyright}
\copyrightyear{2023}
\acmYear{2023}
\acmDOI{XXXXXXX.XXXXXXX}

\acmConference[SIGIR eCom'23]{ACM SIGIR Workshop on eCommerce}{July 27,
  2023}{Taipei, Taiwan}
\acmPrice{15.00}
\acmISBN{978-1-4503-XXXX-X/18/06}

\begin{document}
\sloppy

\title{CDR-Adapter: Learning Adapters to Dig Out More Transferring Ability for Cross-Domain Recommendation Models}

\renewcommand{\thefootnote}{\fnsymbol{footnote}} 

\author{Yanyu Chen$^{1}$\footnotemark[1],
    Yao Yao$^{1}$\footnotemark[1]\footnotetext[1]{Both\ authors\ contributed\ equally\ to\ this\ research.},
    Wai Kin Victor Chan$^{1}$\footnotemark[2]\footnotetext[2]{Corresponding\ author.},
    Li Xiao$^{1}$,
    Kai Zhang$^{2}$,
    Liang Zhang$^{3}$,
    Yun Ye$^{3}$}

 \affiliation{
    \institution{$^{1}$ Tsinghua-Berkeley Shenzhen Institute, Shenzhen International Graduate School, Tsinghua University}
    \city{Shenzhen}
    \state{Guangdong}
    \country{China}
    \postcode{518055}
}
\affiliation{%
  \institution{$^{2}$ Shenzhen International Graduate School, Tsinghua University}
  \city{Shenzhen}
  \state{Guangdong}
  \country{China}
  \postcode{518055}
}
\affiliation{%
  \institution{$^{3}$ Ant Group}
  \city{Shanghai}
  \country{China}
  \postcode{200135}
}

\email{cyy20@tsinghua.org.cn}

\begin{abstract}

Data sparsity and cold-start problems are persistent challenges in recommendation systems. Cross-domain recommendation (CDR) is a promising solution that utilizes knowledge from the source domain to improve the recommendation performance in the target domain. Previous CDR approaches have mainly followed the Embedding and Mapping (EMCDR) framework, which involves learning a mapping function to facilitate knowledge transfer. However, these approaches necessitate re-engineering and re-training the network structure to incorporate transferrable knowledge, which can be computationally expensive and may result in catastrophic forgetting of the original knowledge. In this paper, we present a scalable and efficient paradigm to address data sparsity and cold-start issues in CDR, named \textbf{CDR-Adapter}, by decoupling the original recommendation model from the mapping function, without requiring re-engineering the network structure. Specifically, CDR-Adapter is a novel plug-and-play module that employs adapter modules to align feature representations, allowing for flexible knowledge transfer across different domains and efficient fine-tuning with minimal training costs. We conducted extensive experiments on the benchmark dataset, which demonstrated the effectiveness of our approach over several state-of-the-art CDR approaches.
\end{abstract}

\ccsdesc[500]{Information systems~Recommendation systems}
\ccsdesc[300]{Recommendation systems~Cross-domain recommendation}
\ccsdesc[100]{Computing methodologies~Decoupling}

\keywords{Cross-domain recommendation, Decoupling representation learning, Cold-start problem}

\maketitle

\section{Introduction}

Recommender systems have become increasingly prevalent in online platforms, serving as vital components in providing personalized user experiences~\cite{yao2022razor}. However, a significant challenge in constructing such systems is the provision of accurate recommendations to new users, who possess little or no interaction history with the system, commonly referred to as cold-start users. To overcome this challenge, cross-domain recommendation (CDR) models have emerged as a promising solution that leverages knowledge and information from various domains to enhance recommendation performance.
CDR comprises two domains: the target domain and the source domain, with users in the source domain classified into two groups: overlapping users and cold-start users. Overlapping users have active records in both domains, while the remaining users are considered cold-start users in the target domain. The primary objective of CDR is to boost recommendation performance for cold-start users in the target domain.

\begin{figure*}[ht]
    \centering
    \includegraphics[width=0.95\textwidth]{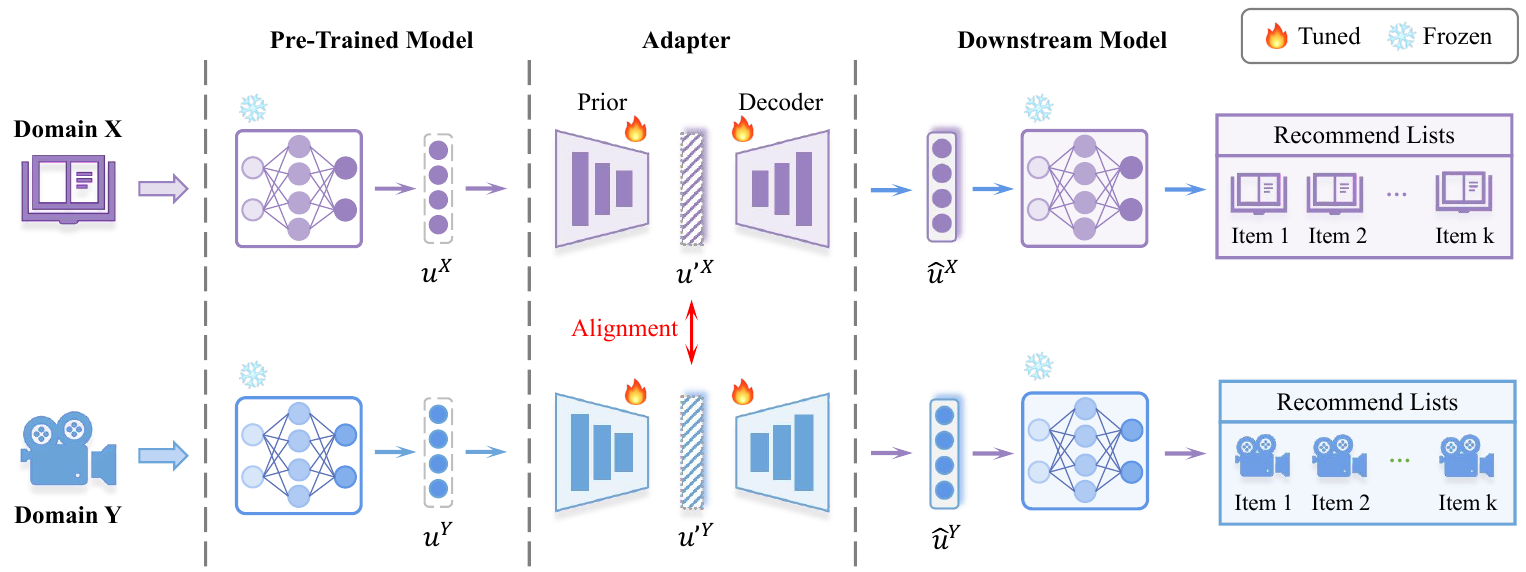}
    \caption{The overall CDR-Adapter framework is composed of three parts: (1) the pre-trained representation models in each domain with fixed parameters; (2) the proposed Adapter trained to align the representations in each domain; (3) the downstream recommendation models with reconstructed alignment representations as input.}
    \label{Adapter_train}
\end{figure*}

Earlier CDR models~\cite{man2017cross, fu2019deeply} mainly focused on learning a cross-domain mapping function that could transfer information and knowledge from the source domain to the target domain, restricted to only the relevant information of the overlapping users, which usually led to sub-optimal recommendation results. Subsequent works~\cite{kang2019semi, zhu2021transfer, zhao2020catn, chen2023clcdr} have improved upon the earlier models by enriching and expanding the transferrable information, such as user-item interaction, thereby reducing the dependence on overlapping users. Despite these advancements, these CDR models continue to have several limitations. Firstly, these models often require a large number of overlapping users to transfer information, which leads to data sparsity issues and models becoming biased toward overlapping users. Secondly, they generally require re-engineering and re-training the network structure to incorporate transferrable information, leading to high computational expenses and risking the catastrophic forgetting of original knowledge. Thirdly, the mapping function learned by earlier models is typically inflexible, which limits the model's ability to transfer knowledge across various domains.

Certain approaches attempt to disentangle domain-specific and cross-domain information. For instance,~\citet{cao2022disencdr} proposed two mutual-information-based disentanglement regularizers that exclusively transfer domain-shared information to enhance model recommendation performance. Additionally,~\citet{cao2022cross} proposed two information bottleneck regularizers to simultaneously model domain-specific and cross-domain information, deriving unbiased representations.  
However, these CDR models necessitate adjusting and retraining the original network to achieve a domain-shared latent space, where representations from different domains are aligned to facilitate knowledge transfer. Training large eCommerce recommendation models can be computationally expensive, and restructuring and retraining the network can alter the intrinsic semantic space of the pre-trained model. Generally, this paradigm suffers from inefficient training and catastrophic forgetting of the original knowledge of pre-trained models.

To address these challenges, we propose a novel cross-domain recommendation framework, CDR-Adapter, inspired by the adapter technique in natural language processing~\cite{he2021effectiveness,ding2023parameter}. Our approach decouples recommendation models from the mapping function by learning an adapter that aligns feature representations of recommendation models across the source and target domains. This preserves the original model information while enabling flexible knowledge transfer. Requiring much less overlapping user information, our approach mitigates the challenges of data sparsity. Scalable and efficient, it allows for efficient fine-tuning with minimal training cost. We evaluate our approach on several benchmark datasets, and the results demonstrate that our method outperforms several state-of-the-art CDR approaches. 

Our main contributions are as follows:
\begin{itemize}
\item We propose a novel CDR framework that leverages adapter modules to align feature representations, enabling flexible knowledge transfer across different domains and efficient fine-tuning with minimal training cost.

\item We introduce a scalable and efficient solution to the cold-start problem in CDR by decoupling recommendation models from the mapping function, without adjusting pre-trained models or facing the problem of catastrophic forgetting.

\item We conduct extensive experiments on several benchmark datasets and demonstrate the effectiveness of our approach over several state-of-the-art CDR approaches.
\end{itemize}

\section{Methodology}

\subsection{Notations and Problem Formulation}
\textbf{Notations.} Without loss of generality, we consider a general CDR scenario where there exist two domains: $X$ and $Y$. 
In this paper, we aim to design an effective and efficient method to improve the recommendation performance in both domains simultaneously, so we do not explicitly differentiate between the source domain and the target domain.
Each domain has its corresponding user set $\mathcal{U}$ and item set $\mathcal{V}$. For simplicity, we further introduce binary matrix $R\in \left\{ 0,1 \right\} ^ {\lvert \mathcal{U} \rvert \times \lvert \mathcal{V} \rvert}$, whose elements indicate whether there is an interaction between a user and an item.

\begin{figure*}[!htbp]
    \centering
    \includegraphics[width=0.95\textwidth]{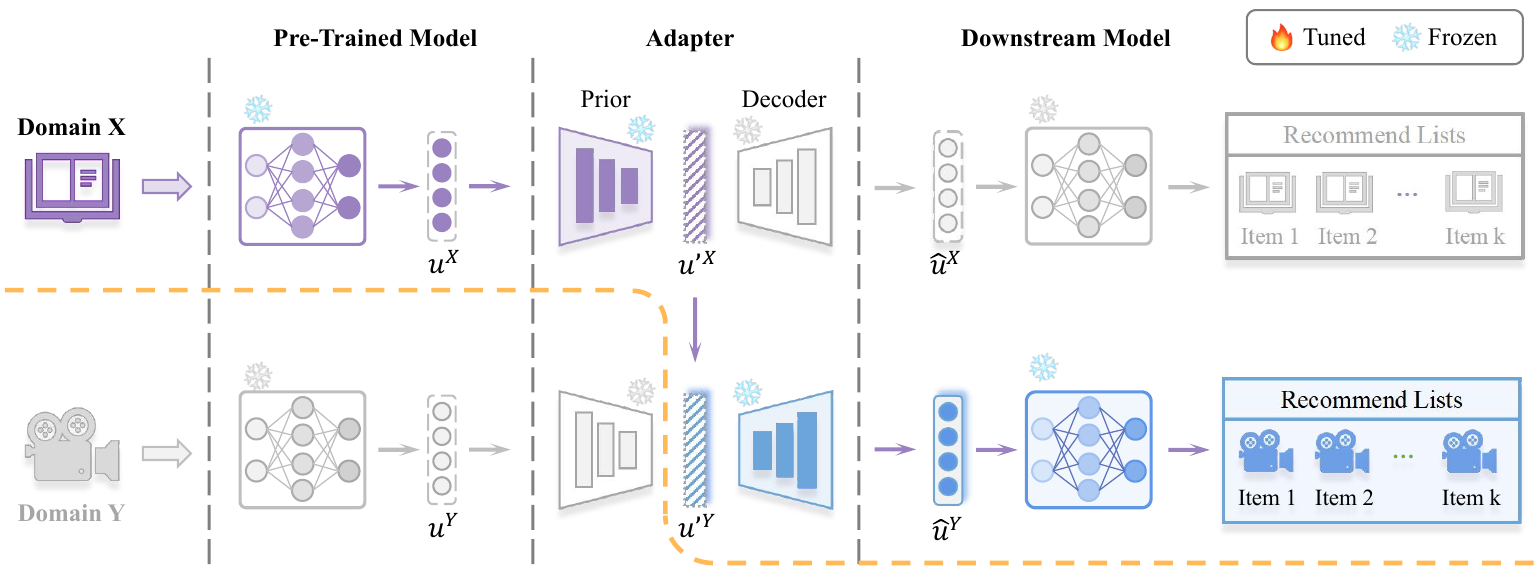}
    \caption{A high-level overview of CDR-Adapter inference process in practice. Above the dotted line, we depict our CDR process for the cold-start user in domain Y: the pre-trained model generates the cold-start user representation in domain X. The representation is fed to the Adapter to produce the accurate representation in domain Y. Then this representation is used as input for the downstream recommendation model to get the final recommendation list. Note that the pre-trained model and downstream model are frozen during the training of the prior and decoder.}
    \label{Adapter_test}
\end{figure*}

\textbf{Problem Formulation.} Given the observed interaction data $R$ of both domains, we aim to make recommendations for cold-start users in domain $Y$, who are only observed in domain $X$ and do not have interaction records in domain $Y$. Formally, given a cold-start user $i\in \mathcal{U}^X$, we would like to recommend a list of items from $\mathcal{V}^Y$ (vice versa, in the case of users from $\mathcal{U}^Y$ and items from $\mathcal{V}^X$).

\subsection{Overview of the CDR-Adapter Framework}

In this paper, we propose to learn a small and simple CDR-Adapter to align the representations generated by two pre-trained models, which can effectively decouple the pre-trained representation model from the downstream recommendation model. The illustration of the components of our method is presented in Figure \ref{Adapter_train}, which is composed of the pre-trained representation models, Adapters and downstream recommendation models. 

Generally, the common recommendation approaches can be divided into two modules: the representation learning module and the downstream recommendation module. The representation learning module is to extract the characteristics of the features and obtains the dense representations in the latent space. The representations learned by large-scale pre-trained representation models have robust generalization properties and can be applied to downstream tasks. The downstream recommendation module is to obtain the recommendation lists for users, which are adaptive to specific scenarios and usually hard to be transferred and generalized.
We start from the representation models, which provide initialized user/item representations (a.k.a. latent variables) in each domain for the following components. Then, the adapter module along with three regularizers, regarded as an auxiliary module, is proposed to align the representations to extract the knowledge across the domain. Specifically, the dimensions of the input and output are aligned, so there are no compatibility issues. Afterward, the alignment representations are reconstructed as input for downstream models to get final recommendation lists for cold-start users.

Figure \ref{Adapter_test} illustrates the inference procedure of CDR-Adapter in practice. Above the dotted line shows the inference process for cold-start users in domain $Y$ by transferring the knowledge from domain $X$. Note that the inference procedure for cold-start users in domain $X$ is similar and omitted for simplicity.
We present an example to further facilitate understanding.
Suppose the task is to obtain a recommendation list for cold-start user A in domain $Y$. We try to transfer the knowledge of user A in domain $X$ to improve the recommendation performance for cold-start user A in domain $Y$. The pre-trained representation model generates the representation of user A in the latent space and feeds representations to the Adapter. The prior is to get the alignment representation for user A in domain $Y$, which can effectively transfer the cross-domain knowledge.
The decoder is to reconstruct the alignment representation in order to fine-tune the input of the downstream model without retraining the model, which fully uses the knowledge in domain $X$ to get the more effective representation and saves amounts of computation cost. Finally, through the downstream recommendation model (e.g. Top-N multi-hop neighborhood recommendation), the recommendation list for cold-start user A can be obtained.

\subsection{Adapter Architecture}
In this paper, we design three tasks as regularizers for the adapter module to capture the correlations across domains, aiming to learn unbiased representations with domain-shared information. Specifically, the contrastive cross-domain regularizer aims to capture the users' correlation across domains. The scale alignment regularizer aims to linearly align the scale of users to map each other in the two domains. The reconstruction information regularizer aims to minimize the loss of information after alignment and reconstruction procedures, which can guarantee that the reconstructed representations can be directly used as input in the downstream recommendation models without retraining the models.

\textbf{Contrastive Cross-Domain Regularizer.}
In order to better align the representations from each domain, we design the contrastive cross-domain regularizer, which improves the capability to make recommendations in both domains. The same overlapping users $(u_{i}^{\prime X}, u_{i}^{\prime Y})$ are regarded as the positive pair. The different users $(u_{i}^{\prime X}, u_{j}^{\prime Y})$ are regarded as the negative pair. We refine the representations of users by measuring the mutual information between the representations from domain $X$ and domain $Y$.
Specifically, the distance between positive pairs is minimized to make the representations aligned in cross-domain, while the distance between negative pairs is maximized to distinguish different users. In this way, the user representations are enforced to capture the domain-shared information from both domains, thus deriving the general representations for CDR. Thus, the contrastive cross-domain regularizer can be formulated as follows.
$$
\mathcal{L}_1=-E\left[ \log \frac{\exp \left[ s\left( u_{i}^{\prime X},u_{i}^{\prime Y} \right) /\tau \right]}{\sum\nolimits_{i,j=1}^N{\left[ s\left( u_{i}^{\prime X},u_{j}^{\prime Y} \right) /\tau \right]}} \right] 
$$

where $u^{\prime}$ is the output alignment representation of the prior layer. $u_{i}^{\prime X}$ and $u_{i}^{\prime Y}$ are the representation of the same overlapping user $i$ in domain $X$ and domain $Y$, while $u_{i}^{\prime X}$ and $u_{j}^{\prime Y}$ are the representation of different overlapping user $i$ in domain $X$ and domain $Y$. $\tau > 0$ is a tunable temperature hyperparameter. $s(u_{i}^{\prime X}, u_{j}^{\prime Y})$ is the cosine similarity between vector $u_{i}^{\prime X}$ and $u_{j}^{\prime Y}$, e.g.,  $s\left( u_i,u_j \right)$ can be calculated as follows:
\begin{equation}
s\left( u_i,u_j \right)=\frac{\left< u_i,u_j \right>}{\left\| u_i \right\| \cdot \left\| u_j \right\|}. \nonumber
\end{equation}

\textbf{Scale Alignment Regularizer.} In order to align the scale of overlapping user representations in each domain, we design a linear scale alignment regularizer to extract domain-shared information to the greatest extent. Ideally, we hope $u^{\prime X} = u^{\prime Y}$, since they are essentially the same users. However, this task is hard to learn and requires high precision of the prior model. So here we propose an approximation method, which is to train the linear transformation which is essentially reciprocal. The formulation is as follows.
$$
F_1\left( u^{\prime X} \right) \,\,=\,\,\alpha _1u^{\prime Y}+\beta _1
$$
$$
F_2\left( u^{\prime Y} \right) \,\,=\,\,\alpha _2u^{\prime X}+\beta _2
$$
Specifically, here we train the Multi-Layer Perceptron (MLP) without the activation layer to obtain the parameters of $\alpha$ and $\beta$. 

Then, the scale alignment regularizer is formulated as follows.
$$
\mathcal{L}_2=\left\| F_1\left( u^{\prime X} \right) -u^{\prime Y} \right\| _2+\left\| F_2\left( u^{\prime Y} \right) -u^{\prime X} \right\| _2
$$

\textbf{Reconstruction Information Regularizer.} To reconstruct the cold-start users' representation for direct prediction in the downstream recommendation models without retraining the downstream model, we propose the reconstructing information regularizer. 
In this part, we hope the reconstructed representation through the decoder can maintain the cross-domain knowledge and has similarity with the original representation, simultaneously.
In this way, the reconstructed representations can be directly used as input for the downstream model to obtain the final recommendation lists. The reconstructing information regularizer is formulated as follows.
$$
\mathcal{L}_3=\left\| u^X-\hat{u}^X \right\| _2+\left\| u^Y-\hat{u}^Y \right\| _2
$$

\textbf{Optimizing the Overall Model.} Based on the above three regularizers, we can optimize the overall model in an end-to-end framework. In summary, we build the prior and decoder to transfer the overlapping users' knowledge. The conclude final objective function is as follows.
$$
\mathcal{L} = \lambda_1 \mathcal{L}_1 + \lambda_2 \mathcal{L}_2 + \lambda_3 \mathcal{L}_3,
$$
where $\lambda_1$, $\lambda_2$, and $\lambda_3$ are the hyper-parameter, which control the importance of each regularizer.

\subsection{The Properties of CDR-Adapter}

The proposed CDR-Adapter can be used to learn alignment representation with a relatively small amount of data, which can perfectly address the challenges of data sparsity and cold-start problems in the recommendation field. 
The CDR-Adapter, as the extra network to inject condition information, also has the following properties.

\textbf{Plug-and-Play.} The original network topology and transfer ability of the existing models does not be changed by adding this CDR-Adapter. Besides, the CDR-Adapter can also be easily composed to any pre-trained and downstream models. There is no need to retrain the pre-trained models and downstream task models.

\textbf{Simple and Small.} It can be easily inserted into any recommendation model with low training costs and fully transfer cross-domain knowledge. It has a small number of parameters and small storage space, which will not introduce much computation cost.

\textbf{Cascade Composable and Flexible.} As to make recommendations in any two domains from the total $n$ domains, the traditional methods need to train $n(n-1)/2$ models, while our method only needs training $n-1$ adapters to make cascaded recommendations. For example, we have three scenarios, \emph{A}, \emph{B}, and \emph{C}. Instead of training three adapters as, \emph{A}$\leftrightarrow$\emph{B}, \emph{A}$\leftrightarrow$\emph{C}, \emph{B}$\leftrightarrow$\emph{C}, it only needs to train two adapters in any two of scenarios, such as \emph{A}$\leftrightarrow$\emph{B}, \emph{B}$\leftrightarrow$\emph{C}. When it refers to making recommendations under the \emph{A}$\leftrightarrow$\emph{C} scenario, we can cascade the two adapters to realize \emph{A}$\leftrightarrow$\emph{B}$\leftrightarrow$\emph{C}.

\textbf{Generalizable.} Once trained, it can be used on custom models as long as they are fine-tuned from the same cross-domain models. No retraining is required for this transfer.

\section{Experiments}
Extensive experiments are conducted to answer the following questions:\\
\textbf{Q1:} How does our CDR-Adapter perform compared with the competitive baselines?\\
\textbf{Q2:} How does the proportion of overlapping users impact the model performance?\\
\textbf{Q3:} Does CDR-Adapter indeed infer more accurate representations of the cold-start users in the latent space? \\
\textbf{Q4:} Does our CDR-Adapter reach the desirable decoupling? Further, what impact does our CDR-Adapter achieve?

\subsection{Experimental Settings}
\textbf{Datasets.} Following previous works\cite{chen2023clcdr}, we adopt the same datasets, and the same preprocessing settings to build our CDR scenarios. Specifically, we conduct experiments on the largest scale of public datasets: Amazon. The most popular pair of domains are selected to evaluate our CDR-Adapters for the bi-directional CDR scenarios. The detailed statistics of each domain are listed in Table \ref{datasets_adapter}.

\begin{table}[htbp]
\centering
\caption{Statistics of the cross-domain scenario. (Overlap. denotes overlapping users.)}
\label{datasets_adapter}
\tabcolsep=0.010\linewidth
\begin{tabular}{cccccc}
	\toprule
	\textbf{Domain}                                     & \textbf{\#Users}                                       & \textbf{\#Items}                                        & \textbf{\#Interactions}                                   & \textbf{Density}                                        & \textbf{\#Overlap.} \\ 
		\midrule
		\begin{tabular}[c]{@{}c@{}}Book\\ Movie\end{tabular}  & \begin{tabular}[c]{@{}c@{}}12,091\\ 8,231\end{tabular} & \begin{tabular}[c]{@{}c@{}}40,122\\ 19,890\end{tabular} & \begin{tabular}[c]{@{}c@{}}583,801\\ 357,656\end{tabular} & \begin{tabular}[c]{@{}c@{}}0.12\%\\ 0.22\%\end{tabular} & 5,644               \\ 
		\bottomrule
	\end{tabular}
\end{table}

\textbf{Implementation Details.} We filter out the items that have fewer than 10 interactions and the users that have fewer than 5 interactions in each domain, making the users/items access to learning effective representation from each domain. We randomly select 20\% as cold-start users for testing and validation (e.g. the 10\% from the Book domain to recommend items in the Movie domain and the residual 10\% from the Movie domain to recommend items in the Book domain) and the remaining users are used for training.

\textbf{Baselines.} In order to verify the effectiveness of our method to cold-start users, we compare our CDR-Adapter with the following state-of-the-art baselines, which can be divided into \emph{three groups}.

\emph{Single-domain recommendation:} The methods in this group consider all domains as a whole single domain. We construct a unified matrix so that it includes all users and items as its rows and columns, respectively. Then, we apply the following widely-used methods.

\begin{itemize}
\item  \textbf{CML}\cite{hsieh2017collaborative} models the user and item representation by matrix learning, which calculates L2 distance and supposes that the distance between a user and interacted items is small while the distance between a user and not interacted items is large. CML is a state-of-the-art collaborative filtering method.

\item \textbf{BPR}\cite{rendle2012bpr} models the latent vector by pairwise ranking loss, which optimizes the order of the inner product of user and item latent vectors.

\item \textbf{NGCF}\cite{wang2019neural} is a graph neural network method to learn user and item representations, which uses GCN to learn high-order information between users and items.
\end{itemize}

\emph{Single-directional cross-domain recommendation:} Single-domain recommendation methods fail to consider the differences between two domains, which makes it hard to effectively transfer knowledge. To better transfer useful knowledge, researchers propose single-direction CDR approaches. We adopt several typical CDR models as baselines. Note that, all of these following methods transfer information from the source to the target domain in one direction, we run two times to achieve bi-directional CDR.

\begin{itemize}
\item \textbf{EMCDR}\cite{man2017cross} first learns user and item representations, and then adopts a network to bridge the representations from the source domain to the target domain.

\item \textbf{SSCDR}\cite{kang2019semi} is a self-supervised bridge-based method that gets the final item list by multi-hop neighborhood inference.

\item \textbf{TMCDR}\cite{zhu2021transfer} is the expansion of EMCDR, which designs a meta-learning framework for CDR to cold-start users.

\item \textbf{CLCDR}\cite{chen2023clcdr} is a contrastive learning-based CDR model, which simultaneously transfers knowledge about overlapping users and user-item interactions to optimize the user and item representations. 

\end{itemize}

\emph{Bi-directional cross-domain recommendation:} Since our method can realize bi-directional CDR, We also compare our method with the following bi-directional CDR methods.

\begin{itemize}
\item \textbf{DAN}\cite{wang2020dual} captures high-order relationships to learn user preferences by utilizing the user-item interaction graph end-to-end.

\item \textbf{DTCDR}\cite{zhu2019dtcdr} designs a kind of multi-task learning to combine the representation of users across the domains and improve the recommendation performance on both richer and sparser domains simultaneously.

\item \textbf{SA-VAE}\cite{salah2021towards} is the state-of-the-art bi-directional CDR method, which is a variational method that utilizes the VAE framework to generate the latent matrix for each domain, and then trains the mapping function for prediction.

\end{itemize}

\textbf{Evaluation Metrics.} Following the previous works\cite{kang2019semi, zhu2021transfer}, we use the leave-one-out evaluation method to verify the effectiveness of our CDR-Adapter. For instance, given a ground truth interaction $(u_i, v_j)$ in domain $Y$, we first randomly select 999 items from the item set $\mathcal{V}^Y$ as negative samples. Then, we calculate 1000 records (1 positive and 999 negative samples) by the learned representation $\hat{u}^Y_i$ from domain $X$ and $v^Y_j$ from domain $Y$. Next, we rank the recommendation list and use evaluation metrics: Hit Rate(HR), Normalized Discounted Cumulative Gain(NDCG), and Mean Reciprocal Rank(MRR) to show the performance of top-N recommendations.

\subsection{Overall Performance (Q1)}

The overall performance is listed in Table \ref{book_movie_over_5}, which reports the mean result under HR, NDCG, and MRR over ten runs with outliers removed. From an overall point of view, our CDR-Adapter method obtains statistically significant improvements compared with the several baselines. We analyze the results from several following perspectives.

\begin{table*}[htbp]
\centering
\caption{Model performance (\%) with 5\% proportion of overlapping users on the bi-directional Book—Movie CDR scenario. The best results are in \textbf{bold-faced} and the best baseline is {\ul underlined}. The last row reports relative improvement over the best baseline. Note that CDR-AP(*) is the abbreviation of our method CDR-Adapter with different pre-trained models. Among them, CDR-AP(HGL) is always statistically significantly better than other models.}
\label{book_movie_over_5}
\tabcolsep=0.0175\linewidth
\begin{tabular}{c|ccccc|ccccc}
\hline
\multirow{3}{*}{\textbf{Method}} & \multicolumn{5}{c|}{\textbf{Book-Domain}}                                                                                                                         & \multicolumn{5}{c}{\textbf{Movie-Domain}}                                                                                                                        \\ \cline{2-11} 
                                 & \multicolumn{2}{c|}{HR}                                          & \multicolumn{2}{c|}{NDCG}                                        & \multirow{2}{*}{MRR}        & \multicolumn{2}{c}{HR}                                           & \multicolumn{2}{c|}{NDCG}                                        & \multirow{2}{*}{MRR}       \\ \cline{2-5} \cline{7-10}
                                 & @10                        & \multicolumn{1}{c|}{@20}            & @10                        & \multicolumn{1}{c|}{@20}            &                             & @10                        & @20                                 & @10                        & \multicolumn{1}{c|}{@20}            &                            \\ \hline
CML                              & 13.32                      & \multicolumn{1}{c|}{18.24}          & 11.31                      & \multicolumn{1}{c|}{14.09}          & 7.23                       & 12.15                      & \multicolumn{1}{c|}{17.34}          & 10.01                      & \multicolumn{1}{c|}{13.12}          & 6.80                      \\
BPR                              & 15.21                      & \multicolumn{1}{c|}{19.34}          & 13.13                      & \multicolumn{1}{c|}{16.08}          & 7.55                       & 14.80                      & \multicolumn{1}{c|}{19.02}          & 12.95                      & \multicolumn{1}{c|}{15.97}          & 7.05                      \\
NGCF                             & 15.28                      & \multicolumn{1}{c|}{20.21}          & 13.42                      & \multicolumn{1}{c|}{16.73}          & 8.02                       & 15.04                      & \multicolumn{1}{c|}{19.67}          & 13.11                      & \multicolumn{1}{c|}{16.25}          & 7.56                      \\ \hline
EMCDR                            & 13.14                      & \multicolumn{1}{c|}{18.52}          & 11.45                      & \multicolumn{1}{c|}{14.99}          & 7.70                       & 15.63                      & \multicolumn{1}{c|}{20.37}          & 13.46                      & \multicolumn{1}{c|}{16.83}          & 8.11                      \\
SSCDR                            & 16.26                      & \multicolumn{1}{c|}{21.86}          & 14.90                      & \multicolumn{1}{c|}{17.57}          & 9.23                       & 18.49                      & \multicolumn{1}{c|}{22.69}          & 15.77                      & \multicolumn{1}{c|}{18.70}          & 18.03                      \\
TMCDR                            & 19.64                      & \multicolumn{1}{c|}{22.64}          & 15.77                      & \multicolumn{1}{c|}{18.51}          & 8.62                       & 20.10                      & \multicolumn{1}{c|}{23.73}          & 16.89                      & \multicolumn{1}{c|}{19.62}          & 9.66                      \\
CLCDR                            & 20.83                      & \multicolumn{1}{c|}{{\ul 23.76}}    & 17.69                      & \multicolumn{1}{c|}{20.95}          & {\ul 9.28}                       & {\ul 21.35}                & \multicolumn{1}{c|}{{\ul 24.41}}    & 18.40               & \multicolumn{1}{c|}{{\ul 21.07}}    & {\ul 10.15}                \\ \hline
DAN                              & 18.01                      & \multicolumn{1}{c|}{22.65}          & 15.78                      & \multicolumn{1}{c|}{18.55}          & 8.34                       & 19.24                      & \multicolumn{1}{c|}{23.44}          & 16.20                      & \multicolumn{1}{c|}{19.21}          & 9.57                      \\
DTCDR                            & 20.69                      & \multicolumn{1}{c|}{23.11}          & 16.42                      & \multicolumn{1}{c|}{19.70}          & 8.83                       & 20.13                      & \multicolumn{1}{c|}{23.80}          & 17.01                      & \multicolumn{1}{c|}{19.94}          & 9.27                      \\
SA-VAE                            & {\ul 21.79}                & \multicolumn{1}{c|}{23.09}          & {\ul 18.63}                & \multicolumn{1}{c|}{{\ul 21.03}}    & 9.02                 & 21.04                      & \multicolumn{1}{c|}{23.25}          & {\ul 18.96}                      & \multicolumn{1}{c|}{21.05}          & 9.67                      \\ \hline
\textbf{CDR-AP(BPR)}              & 21.88                      & \multicolumn{1}{c|}{23.14}          & 18.42                      & \multicolumn{1}{c|}{21.69}          & 11.53                       & 22.15                      & \multicolumn{1}{c|}{23.77}          & 19.38                      & \multicolumn{1}{c|}{22.01}          & 11.58                      \\
\textbf{CDR-AP(NGCF)}              & 24.31                      & \multicolumn{1}{c|}{27.29}          & 21.72                      & \multicolumn{1}{c|}{24.15}          & 11.59                       & 24.83                      & \multicolumn{1}{c|}{27.08}          & 21.64                      & \multicolumn{1}{c|}{23.79}          & 11.65                      \\
\textbf{CDR-AP(HGL)}             & \textbf{25.22*}             & \multicolumn{1}{c|}{\textbf{27.98*}} & \textbf{22.67*}             & \multicolumn{1}{c|}{\textbf{24.89*}} & \textbf{12.03*}              & \textbf{25.01*}             & \multicolumn{1}{c|}{\textbf{27.85*}} & \textbf{23.42*}             & \multicolumn{1}{c|}{\textbf{25.33*}} & \textbf{11.87*}             \\ \hline
\textbf{Improvement}             & \multicolumn{1}{c}{15.7\%} & \multicolumn{1}{c|}{17.8\%}         & \multicolumn{1}{c}{21.7\%} & \multicolumn{1}{c|}{18.4\%}         & \multicolumn{1}{c|}{29.6\%} & \multicolumn{1}{c}{17.1\%} & \multicolumn{1}{c|}{14.1\%}         & \multicolumn{1}{c}{23.5\%} & \multicolumn{1}{c|}{20.2\%}         & \multicolumn{1}{c}{16.9\%} \\ \hline
\end{tabular}%
\vspace{1mm}
$*$ indicates that the improvements are statistically significant for p < 0.05 judged with the runner-up result in each case by paired t-test.
\end{table*}

\vspace{2mm}
\textbf{Comparison with Single-Domain Models.} (1) First, we found that, the graph-based model NGCF consistently outperform CML and BPR in term of all evaluation metrics, which indicates that transferring the multi-hop neighborhood knowledge is effective for learning better user and item representations. 
(2) Second, the performance of our CDR-Adapter with different pre-trained models demonstrates the significant importance of user and item representations generated by pre-trained models. The final performance of the CDR-Adapter model is positively correlated with the domain-specific representation generated by the pre-trained model. 
(3) Third, these single-domain models perform mostly worse than the cross-domain models, due to these methods ignoring the difference between different domains just combining them together in a simple way and making recommendations, which is hard to learn the transferable knowledge for cold-start users. So it is necessary to dig out transferring ability for CDR to cold-start users.

\textbf{Comparison with Single-Directional Cross-Domain Recommendation Models.} (1) First, in general, the cross-domain methods are superior to corresponding single-domain methods, which demonstrates that adopting different transferring components for CDR is better than using one single neural network to model the mixed matrix.
(2) Second, the improvement of the EMCDR model is limited and even worse than some single-domain models, which indicates that a simple function may be not effective to learn the complex mapping relation of cross-domain representations. 
(3) Third, since the single-directional CDR models could only improve the recommendation performance in the target domain, it should be run two times to achieve bi-directional CDR, which requires high computing costs and time consumption. Besides, the transfer might be negative transferring in some cases.
(4) Forth, since EMCDR-based models mainly transfer the overlapping users' information, user-item interactions, and even user-user social relationships, the generative representations would be biased toward overlapping users. Compared with all EMCDR-based baselines, our CDR-Adapter achieves statistically significant improvements with all evaluation metrics, which demonstrates that learning the mapping function on the biased representations can be hard to obtain the optimal results. In contrast, our CDR-Adapter digs out the transferring ability of domains and utilizes three kinds of regularizers to encourage the representations to focus on aligning the domain-specific representations. In this way, the unbiased cold-start user representation on each domain can be directly obtained in the target domain.

\textbf{Comparison with Bi-Directional Cross-Domain Recommendation Models.} (1) The recommendation results of DTCDR and SA-VAE in two domains improve, due to their dual objective optimization. (2) Since those bi-directional CDR models jointly optimize the objective by overlapping users, the existing models are still not capable of effectively capturing the domain-shared information in the small number of overlapping users case. While our CDR-Adapter can achieve better performance by aligning the representation of both overlapping users and domain-specific users.

\subsection{The Impact of Overlapping Users (Q2)}
\begin{figure*}[htbp]%
    \centering
    \includegraphics[width=1\textwidth]{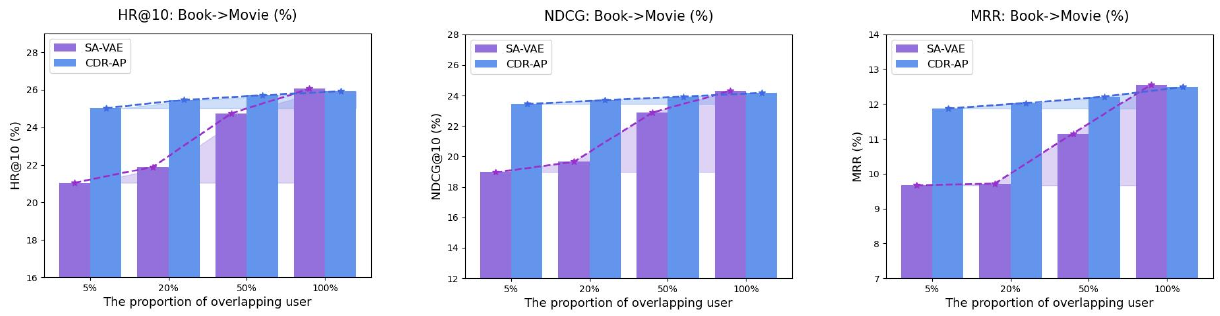}
    \caption{The impact of the proportion of overlapping users in three scenarios.}
    \label{overlap_user_impact}
\end{figure*}

To study the robustness of our CDR-Adapter method regarding the proportion of overlapping users, we conduct several experiments with a certain proportion  $\eta \in \left\{ 5\%, 20\%, 50\%, 100\% \right\} $ of overlapping users for training. Table \ref{overlapping user_proportion} and Figure \ref{overlap_user_impact} show the recommendation performances of SA-VAE and CDR-Adapter on the cross-domain scenario (e.g., the target Movie domain with the source Book domain).

\begin{table}[ht]
\centering
\caption{The robustness performance (\%). The $\eta$ denotes the proportion of overlapping users in the training process.}
\label{overlapping user_proportion}
\tabcolsep=0.014\linewidth
\begin{tabular}{c|ccc|ccc}
\hline
\multirow{2}{*}{$\eta$} & \multicolumn{3}{c|}{SA-VAE}                  & \multicolumn{3}{c}{\textbf{CDR-AP}}                            \\ \cline{2-7} 
                   & HR@10 & NDCG@10  & MRR & HR@10 & NDCG@10  & MRR \\ \hline
5\%                & 21.04      & 18.96       & 9.67     & 25.01    & 23.42      & 11.87   \\
 20\%               & 21.89      & 19.65       & 9.72     & 25.46    & 23.70      & 12.03   \\
  50\%               & 24.73      & 22.86       & 11.16      & 25.71    & 23.93      & 12.22   \\
 100\%              & 26.05      & 24.31       & 12.56      & 25.92    & 24.17      & 12.49   \\  \hline
\end{tabular}%
\end{table}

From Table \ref{overlapping user_proportion}, we have the following findings. (1) With the proportion of overlapping users increasing, the performance of SA-VAE steadily improves, which relies on transferring overlapping users to enhance the correlation across the domains. While CDR-Adapter is not sensitive to the proportion of overlapping users. (2) Compare with the strongest baseline, CDR-Adapter performs well even in the 5\% overlapping users, which shows strong robustness. The results reveal that our CDR-Adapter is capable of effectively mapping even in the small number of overlapping users in the training process, due to its useful alignment function, which successfully overcomes the limitation of the existing methods. From Figure \ref{overlap_user_impact}, we find that CDR-Adapter is robust enough, especially in the data sparsity domain (e.g., Movie). In contrast, the improvement is not obvious (e.g., Book). In addition, in the case of a few overlapping users (e.g., 5\%), our CDR-Adapter method achieves nearly the same performance as the SA-VAE in the case of 100\% overlapping users.

\subsection{The Analysis of Latent Space Inference (Q3)}
To further investigate why our CDR-Adapter outperforms the state-of-the-art method, we conduct both qualitative and quantitative analyses on the target domain latent space. 

\begin{figure*}[htbp]%
    \centering
    \includegraphics[width=1\textwidth]{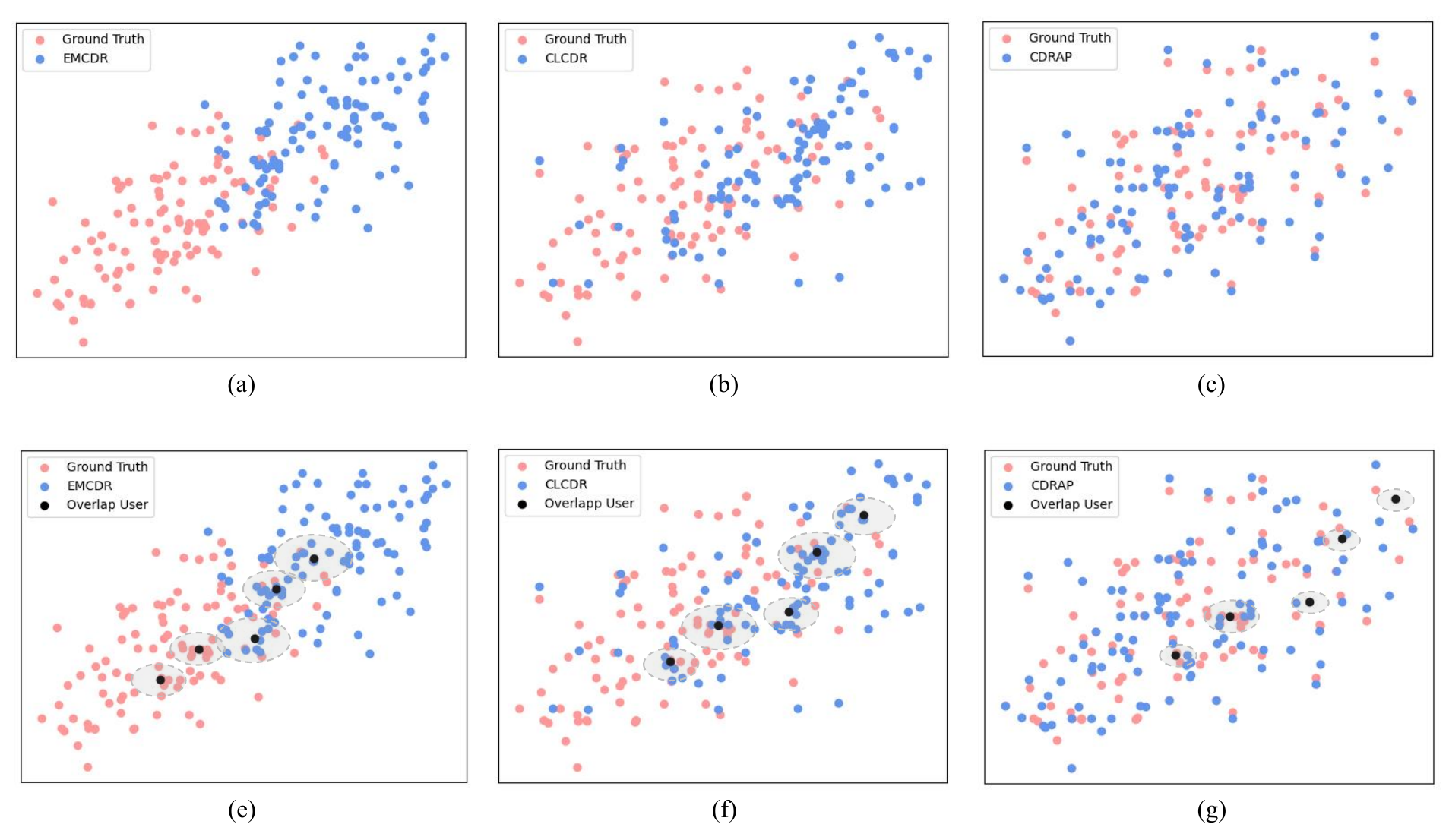}
    \caption{The t-SNE visualization of transferred user representations on Book$\rightarrow$Movie scenario. The inferred user representations by EMCDR, CLCDR, and CDR-Adapter are shown with blue dots and the ground truths are shown with pink dots. The black dots and translucent circles denote the overlapping users and their surrounding areas, respectively.}
    \label{t_sne}
\end{figure*}

As for the qualitative analysis, we make the target domain latent space visualization by adopting t-distributed Stochastic Neighbor Embedding\cite{van2008visualizing} (t-SNE) to analyze the transferring quality compared with ground truth. As Figure \ref{t_sne} shows, the first row is the user representation in the target domain latent space. The inferred user representations by EMCDR, CLCDR, and CDR-Adapter are shown with blue dots and the ground truths are shown with pink dots. Because the EMCDR method infers user representation just by transferring the overlapping user information, the inferred representations are not very close to ground truths. The CLCDR method transfers the information of overlapping users and user-item interactions and designs two contrastive-based domain-specific and domain-shared loss functions, which can improve the representation quality, shown in Figure \ref{t_sne}(b). Our method CDR-Adapter aligns the domain-specific representation instead of mapping and changing the latent space, which can infer the cold-start user representation close to the ground truths.

The second row in Figure \ref{t_sne} is the gathering situation. The black dots and translucent circles denote the overlapping users and their surrounding areas, respectively. The user representation obtained by EMCDR and CLCDR inference is biased toward overlapping users. We can find that there are more users clustered around overlapping users, because these two methods improve the cold-start user recommendation effect mainly by transferring overlapping users' information. While our CDR-Adapter can infer the less biased cold-start user representation in the target domain latent space.

For the quantitative analysis, we measure the actual average distance between the inferred latent representation and their ground truths, shown in Table \ref{mean_distance}. Our CDR-Adapter performs best with the smallest distance.

\begin{table}[htbp]
\centering
\caption{The average distance between the inferred user representations and ground truths.}
\label{mean_distance}
\tabcolsep=0.020\linewidth
\begin{tabular}{c|cccc}
\toprule
Method           & EMCDR & CLCDR & SA-VAE & CDR-AP \\ \midrule
Average Distance & 1.05  & 0.87  & 0.80   & 0.27   \\ \bottomrule
\end{tabular}%
\end{table}

\subsection{The Analysis of Disentanglement (Q4)}
Table \ref{KL_divergence} demonstrates that our CDR-Adapter method achieves the desirable disentanglement to learn domain-specific and cross-domain representations for users. Specifically, we calculate the average KL divergence to measure the mutual information of domain-specific user representation and cross-domain user representation after the training process of both domains is finished (higher KL divergence means lower mutual information). According to Table \ref{KL_divergence}, it is obvious that our CDR-Adapter method KL divergence is much higher than the EMCDR method, which verifies our CDR-Adapter achieves the outstanding disentanglement between domain-specific and cross-domain representations.

\begin{table}[htbp]
\centering
\caption{The KL divergence between domain-specific and cross-domain representations.}
\label{KL_divergence}
\tabcolsep=0.08\columnwidth
\begin{tabular}{cc}
\toprule
Method      & Scenario: Book$\rightarrow$Movie  \\ \midrule
EMCDR       & 40.26                                    \\
CDR-AP & 273.72                          \\ \bottomrule
\end{tabular}%
\end{table}

\section{Conclusion}
In summary, we proposed a novel and scalable cross-domain recommendation paradigm that addresses the limitations of existing approaches by introducing an adapter plugin that decouples the original representation model from the mapping function. Our approach can align the feature representations of the recommendation models in both domains and enable efficient finetuning with minimal training cost. It can be easily plugged in between the pre-train representation module and the downstream recommendation module of any kind of cross-domain recommendation model.
We believe that our framework provides a more scalable and efficient solution to the cold-start problem in the cross-domain recommendation and offers a promising direction for future research.

\begin{acks}
This research was funded by the Ant Group through CCF-Ant Research Fund (CCF-AFSG RF20220216), the Science and Technology Innovation Commission of Shenzhen (JCYJ20210324135011030), Guangdong Pearl River Plan (2019QN01X890), and National Natural Science Foundation of China (Grant No. 71971127).
\end{acks}

\bibliographystyle{ACM-Reference-Format}
\bibliography{sample-base}

\end{document}